\documentclass[graybox]{svmult}

\usepackage{mathptmx}       
\usepackage{helvet}         
\usepackage{courier}        
\usepackage{type1cm}        
\usepackage{makeidx}         
\usepackage{graphicx}        
\usepackage{multicol}        
\usepackage[bottom]{footmisc}
\usepackage{amsmath}
\usepackage{units}
\usepackage{cases}
\usepackage{xcolor}
 \usepackage{ulem}

\makeindex             

\begin{document}

\title*{Automated Quality Assessment of Space-Continuous Models for Pedestrian Dynamics}
\author{Valentina Kurtc, Mohcine Chraibi and Antoine Tordeux}
\institute{Valentina Kurtc \at Peter the Great St. Petersburg Polytechnic University, Polytechnicheskaya, 29, St. Petersburg, Russia, \email{kurtsvv@gmail.com}
\and Mohcine Chraibi \at Forschungszentrum J\"ulich \email{m.chraibi@fz-juelich.de}
\and Antoine Tordeux \at University of Wuppertal, \email{tordeux@uni-wuppertal.de}}
%
%
\maketitle


 \abstract{printed version.}
 In this work we propose a methodology for assessment of pedestrian models continuous in space.
 With respect to the Kolmogorov-Smirnov distance between two data clouds, representing for instance
 simulated and the corresponding empirical data, we calculate an evaluation factor between zero and one. 
 Based on the value of the herein developed factor, we make a statement about the goodness of the model under evaluation.
 Moreover this process can be repeated in an automatic way in order to maximize the above mentioned factor and hence determine the optimal set of model parameters.

\section{Introduction}
Mathematical models have been developed to describe and simulate the dynamics of complex systems. For pedestrian dynamics several kinds of models have been built with different intentions. For an overview of the existing models the reader is referred to \cite{Schadschneider2009a}. In order to develop and use mathematical models for pedestrian dynamics with high fidelity level, the verification and validation process should be considered as key part in the development cycle.

According to conventional terminology \cite{Sargent2013} \textbf{verification} is the process of assuring if the computer programming and implementation of a mathematical model is correct.
The verification process does not assess the quality of mathematical models nor does it allow to make any conclusions related to its realism.
In the literature some verification-driven works regarding the goodness-of-fit of a pedestrian model have been published. In \cite{IMO2007}
 several verification tests were recommended to assess the quality of simulations of ships. In the RiMEA project \cite{RIMEA2007} verification tests were proposed to evaluate the quality of pedestrian simulations.
Based on the above-mentioned works Ronchi et al. \cite{Ronchi2013a} suggested some modified and new verification tests with the aim to additionally consider the movement of pedestrians in case of fire.

Having verified the numerical implementation of the given model, the question that should be of concern is ``does the implemented model show sufficient accuracy in emulating the targeted system?'' In order to answer this question reliably, an iterative \textbf{validation} process is required, that is, comparing the model/simulation results with empirical findings and evaluating their discrepancies to enhance the behavior of the model. Repeating this process leads to a better description of the targeted system (pedestrian dynamics). The process of validation in pedestrian dynamics can benefit from the tremendous development of empirical research. Several experiments were  conducted to investigate crowd performances and to build a well-documented database. Empirical research does not only contain data of controlled experiments  \cite{Hoogendoorn2005,Kretz2006,Kretz2006a,LiuX2009,Chattaraj2009,Moussaid2009a,Boltes2011,Jelic2012,Holl2013,Burghardt2014}, but also data issued from field studies \cite{Burghardt2010,Burghardt2013a,Seer2014}. These empirical works capture relevant properties of the crowd in \textit{normal} situations and for basic geometries. 
Examples are unidirectional flows (fundamental diagram), bottleneck (jam formation), counter-flow (lane formation), intersection (e.g. T-junction).

The goal of this paper is to develop an automatic assessment of the predictive capability of pedestrian models. We introduce new quantities, in order to quantify the degree of success (or failure) of a model/simulation with respect to a given set of empirical data. The introduced ``validity factor'' is a measure for the verification \textit{and} the validation process of a given implementation of a model. In the first stage of the model's assessment (verification)  we use tests mostly based on the RiMEA guideline \cite{RIMEA2007}. While in the second stage (validation) we compare the model with experimental data issued from six different experiments: Uni-directional flow (1D and 2D), bi-directional flow, corners, bottlenecks and T-junctions.  Comparing the model results with empirical findings and automatically evaluating their discrepancies is important for the validation process as well as for making a precise estimation of its goodness.
1\section{Verification and Validation Tests}
\begin{figure}[h]
\centering{\includegraphics[scale=0.6]{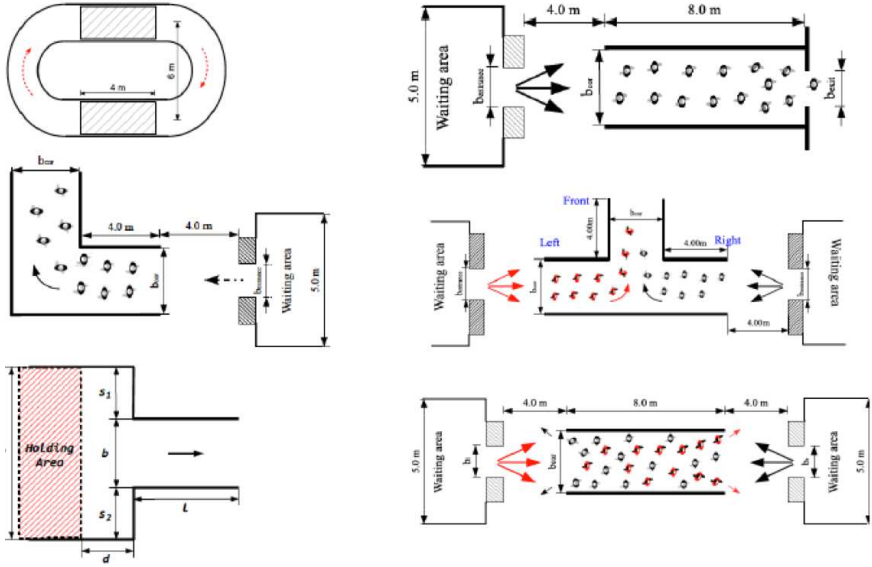}}
\caption{Validation tests: six different experimental set-ups suggested to validate a model.}
\label{fig:ValTests}       
\end{figure}

We implement two different kinds of tests: verification and validation tests. In the first stage of the pedestrian model assessment, the verification step, we use tests based on the RiMEA guideline~\cite{RIMEA2007}.
During the validation process we propose to compare the simulation results with experimental data issued from six different experiments: Uni-directional flow (1D and 2D), bi-directional flow, corners, bottlenecks and T-junctions (Fig.~\ref{fig:ValTests}). 
The proposed geometries are different in order to cover various dynamics of pedestrians and hence, the validation results may show more significance with respect to empirical findings.
  
Considering the pedestrian model and specifying area geometry and initial conditions, which are speed and position of pedestrians, one can compute trajectories of all pedestrians. 
The simulations are performed with JuPedSim \cite{KemlohWagoum2015,jupedsim}, using models \cite{Tordeux2016,Tordeux2016b}. 

Afterwards, both for simulated and empirical trajectories the fundamental diagram is calculated using one of the measurement methods implemented in the JPSreport module of JuPedSim. For an accurate comparison it is important to use the same measurement method for both the experimental as well as the simulation data.

\section{Methodology}
This section presents the methodology for quantitative comparison of results obtained from experiments and simulations on basis of the verification and validation tests introduced earlier. 
Firstly, we suggest the approach for similarity assessment of two fundamental diagrams (speed-density relations), which are the main outcome of the validation tests. 
This approach exploits the data-binning and cumulative distribution functions calculation. 
Secondly, the quantitative quality assessment metric is formulated, which aggregates the comparison results of experiments and simulations both for verification and validation tests.
\subsection{Comparison of two data clouds}
\label{subsec:2DataCloudsComp}
\begin{figure}[h]
\begin{center}
\includegraphics[width=\textwidth]{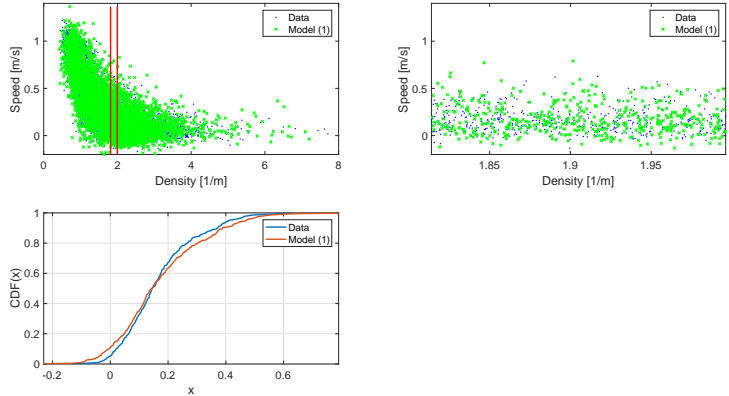}
\caption{Comparison of simulated (green) and empirical (blue) speed-density relations with data-binning method (upper plots). Cumulative distribution functions for specific bin (bottom plot).}
\label{fig:DataBin}       
\end{center}
\end{figure}

Assessment of validation results is made by means of comparison of two fundamental diagrams (FD), i.e. the speed-density relation. 
Given observation points of speed and density from experiments  $\{[\rho_i^{\text{data}},v_i^{\text{data}}]\}$ and simulations $\{[\rho_i^{\text {model}},v_i^{\text{model}}]\}$, our goal is to quantify the degree of similarity 
among these two point clouds. 
Therefore, we create a partitioning of the data points by filtering them according to $N$ equally-spaced density intervals  (Fig.~\ref{fig:DataBin}),
\begin{equation}
	V_j^{\text{src}}=\{v_i^{\text{src}}:\rho_i^{\text{src}} \in [\rho_j,\rho_{j+1}],i=1,...,N^{\text{src}}\},
\label{eq:09}
\end{equation}
where $j=1,...,N, \text{src}=\{\text{data,\, model}\}$ and $N^{\text{src}}$ is the number of observations. 
The key idea here is
to interpret $V_j^{\text{src}}$ as a one-dimensional random variable. Now
we calculate the cumulative distribution functions (CDFs) both for the experiment,
{$F_{V_j^{\text{data}}}(x)$} as well as for the simulations, {$F_{V_j^{\text {model}}}(x)$},
and calculate the Kolmogorov-Smirnov distance
\begin{equation}
	D_j=\sup_{x}|F_{V_j^{\text{data}}}(x)-F_{V_j^{\text{model}}}(x)|,j=1,...,N.
\label{eq:10}
\end{equation}
Finally, the weighted arithmetic mean of distances $D_j$  is used as a quantitative metric estimating the similarity of two data-clouds
\begin{equation}
	D^*=\frac{\sum_{j=1}^N (N_j^{\text{data}}+N_j^{\text{model}})D_j}{N^{\text{data}}+N^{\text{mode}l}},
\label{eq:11}
\end{equation}
where $N_j^{\text{data}}$ and $N_j^{\text{model}}$ are the number of observations in the $j$th bin for data and model respectively. In other words, metric $D^*$ quantifies the degree of success (or failure) of the validation process.

\subsection{Validity factor}

Verification tests are considered to be relatively simple and are expected to be fulfilled. 
For each verification test there are only two possible outputs which are 0 (failure) and 1 (success). 
The quantitative metric for $N_{\text{ver}}$ verification tests is
\begin{equation}
	\delta_{\text{ver}}=\displaystyle\prod_{k=1}^{N_{\text{ver}}} v^{\text{ver}}_k,
\label{eq:03}
\end{equation}
where $v^{\text{ver}}_k$ is the outcome of the $k$th verification test. 

We quantify the degree of success of the considered pedestrian model for the $k$-th validation test with $v^{\text{val}}_k\in[0,1]$.  0 and 1 corresponds to complete failure and absolute success respectively. 
The value $v^{\text{val}}_k$ is computed according to the methodology described in Section~\ref{subsec:2DataCloudsComp}, that is $v^{\text{val}}_k=1-D^*$. 
To quantify the degree of success for $N_{\text{val}}$ validation tests we  write
\begin{equation}
	\delta_{\text{val}}=\frac{\displaystyle\sum_{k=1}^{N_{\text{val}}} {w_k}\cdot v^{\text{val}}_k}{N_{\text{val}}},
\label{eq:01}
\end{equation}
with a weight $w_k$ equal to 1. 

However, in general the parameter values of the model are obtained after a calibration process involving  several scenarios simultaneously \cite{Campanella2014}. 
In this case we calculate $w_k$ as follows
\begin{equation}
	w_k=\frac{N_{\text{cal}}}{N_{\text{val}}},
\label{eq:02}
\end{equation}
where $N_{\text{cal}}\in[0,N_{\text{val}}]$ is a number of scenarios used simultaneously in the calibration procedure. 
Eq. (\ref{eq:02}) expresses the following assumption: it is less beneficial to get the value $v^{\text{val}}_k$ by calibration obtained on the basis of one scenario, that is $w_k=\frac{1}{N_{\text{val}}}$, than if we get the same value $v^{\text{val}}_k$ with the parameter set from calibration with all scenarios simultaneously ($w_k=1$).

Finally the success estimator of the considered model on the basis of verification and validation tests is as follows
\begin{equation}
	\delta=\delta_{\text{ver}}\times{\delta_{\text{val}}}.
\label{eq:04}
\end{equation}

\section{Results}
\label{sec:results}
\begin{figure}[h]
\centering{\includegraphics[width=\textwidth]{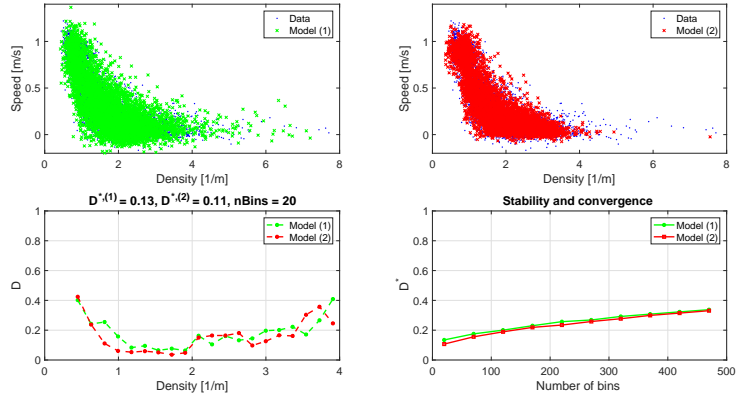}}
\caption{Uni-directional 1D flow. Fundamental diagrams calculated on the basis of real trajectories and trajectories produced with Model (1) (upper-left) and Model (2) (upper-right). The Kolmogorov-Smirnov distance (\ref{eq:10}) for each density bin (lower-left). The dependence of the value $D^*$ (\ref{eq:11}) on the number of density bins $N$.}
\label{fig:FDsComparison}       
\end{figure}

We tested this methodology on two continuous pedestrian speed models based on optimal velocity function and specific stochastic additive noises. 
The noise is white for the first model (Model (1)) \cite{Tordeux2016}, while it is determined by the inertial Ornstein-Uhlenbeck process for the second model (Model (2)) \cite{Tordeux2016b}. 
In this section we show only results of one of the validation tests -- uni-directional 1D flow. 
In other validation test (except bottleneck scenario) the fundamental diagram is produced by the measurement module (Jpsreport) using  Voronoi diagrams (method D). 
Figure \ref{fig:FDsComparison} shows the results. 
The first row represents speed-density relations calculated on the basis of real trajectories and obtained from simulations of Model (1) and Model (2). 

We consider 20 bins for the density partitioning of the validation test and limit the observations for densities not higher than $4\, m^{-2}$. 
For both models and each density bin the Kolmogorov-Smirnov distance is calculated (lower-left subplot  of Fig.\ \ref{fig:FDsComparison}), the mean value of which (Eq. \ref{eq:11}) is equal to $0.13$ and $0.11$ for Model (1) and Model (2) respectively. 

Moreover, we investigated how the number of density bins $N$ influences the final value $D^*$. 
Considering values of $N$ from $20$ to $470$ we calculated corresponding values of $D^*$ (bottom-right subplot of Fig.\ \ref{fig:FDsComparison}). 
Firstly, one can conclude that Model (2) is always better than Model (1), which is conform with the results in \cite{Tordeux2016b}. 
Secondly, for both models the value of $D^*$ is increasing when the number of bins $N$ is increasing. 
This can be explained by the fact that, with increasing number of bins, the number of points each bin contains, becomes more scarce. 
For example, if a bin contains only two points from each compared data set it is obvious that the Kolmogorov-Smirnov distance (Eq. \ref{eq:10}) will be large. 

\section{Discussion and Conclusions}

In this paper we present a methodology for quantitative assessment of pedestrian models exploiting results of verification and validation tests. 
However, this procedure can be applied to any space-continuous model describing pedestrian dynamics. 
Two stochastic pedestrian models were compared using the results of 1D uni-directional flow test. 
We investigated stability properties of the proposed metric (\ref{eq:11}) by considering different number of bins for density partitioning. 
According to the results, the metric for quantitative assessment shows a monotonic increasing behavior.

More generally, the method of quantitative comparison of two data-clouds (here, two fundamental diagrams) 
using an averaged Kolmogorov-Smirnov distance $D^*$ (Section ~\ref{sec:results}) can be interpreted as a separate result. In other words, this method can be used for comparison of two data-clouds of any nature.

The approach for comparison of two data-clouds allows a detailed analysis for different density ranges (lower densities, higher densities) by means of Kolmogorov-Smirnov distances per density interval $D_j$, $j=1,...,N$. 
Analyzing values of $D_j$, it is possible to determine for which density ranges the underlying model performs better wrt. to experimental data.

Finally, the proposed metric (\ref{eq:11}) can be used in calibration procedure as the minimized objective function. 
We suggest to use the introduced quantity $D^*$ (\ref{eq:11}) as a goodness of fit function in calibration procedure.




\bibliographystyle{spmpsci}
\bibliography{ped}
\end{document}